\def\paperversiondraft{draft}

\ifx\paperversion\paperversiondraft
\else
	\ifx\paperversion\paperversioncameraACM
	\else
		\def\paperversion{normal}
	\fi
\fi

\documentclass[a4paper,USenglish,cleveref, autoref, thm-restate]{lipics-v2021}

\usepackage{subcaption}
\usepackage{syntax}
\usepackage{colortbl}

\usepackage{environ}
\usepackage{wasysym}
\usepackage{xspace}
\ifx\paperversion\paperversiondraft
	
\else
	\NewEnviron{draftonly}{}
\fi

\usepackage[utf8]{inputenc}
\usepackage[T1]{fontenc}
\usepackage{microtype}

\usepackage{supertabular} 
\usepackage[frozencache,cachedir=.]{minted}
\usepackage{mdframed}
\newminted[mlir]{text}{mathescape, fontsize=\footnotesize}
\newminted[llvm]{text}{fontsize=\footnotesize, bgcolor=LightGrayColor}
\newminted[ccode]{text}{fontsize=\footnotesize, bgcolor=LightGrayColor}
\newminted[lean]{lean}{fontsize=\small, escapeinside=@@, mathescape}
\newminted[leanfootnotesize]{lean}{fontsize=\footnotesize, escapeinside=@@, mathescape}

\BeforeBeginEnvironment{mlir}{\begin{mdframed}[linewidth=2,topline=false,rightline=false,bottomline=false,skipabove=.5em,linecolor=black!20, innerleftmargin=.7em]}
		\AfterEndEnvironment{mlir}{\end{mdframed}\vspace{.5em}}

\usepackage{xcolor}
\usepackage{xargs}
\usepackage{lipsum}
\usepackage{xparse}
\usepackage{xifthen, xstring}
\usepackage{xspace}
\usepackage{marginnote}
\usepackage{etoolbox}
\usepackage{amssymb}
\usepackage{amsmath}
\usepackage{stmaryrd}

\usepackage[textsize=tiny]{todonotes}
\usepackage[normalem]{ulem}

\makeatletter
\font\uwavefont=lasyb10 scaled 652
\newcommand\colorwave[1][blue]{\bgroup\markoverwith{\lower3\p@\hbox{\uwavefont\textcolor{#1}{\char58}}}\ULon}

\newcommand\highlight[2]{{\color{#1}{\colorwave[#1]{#2}}}}
\makeatother

\makeatletter
\newcommand\InFloat[2]{\ifnum\@floatpenalty<0\relax#1\else#2\fi}
\makeatother

\ifx\paperversion\paperversiondraft
\newcommand\createtodoauthor[2]{
  \def\tmpdefault{emptystring}
  \expandafter\newcommand\csname #1\endcsname[2][\tmpdefault]{
    \def\tmp{##1}
    \InFloat{
        \smash{
	  \marginnote{
	    \todo[inline,linecolor=#2,backgroundcolor=#2,bordercolor=#2]
	      {\textbf{#1 (Figure):} ##2}
          }
        }
    }{\ifthenelse{\equal{\tmp}{\tmpdefault}} 
      {\todo[linecolor=#2,backgroundcolor=#2,bordercolor=#2]{\textbf{#1:} ##2}\ignorespaces}
      {\ifthenelse{\equal{##2}{}} 
        {\highlight{#2}{##1}}
        {\highlight{#2}{##1}\todo[linecolor=#2,backgroundcolor=#2,bordercolor=#2]
	  {\textbf{#1:} ##2}
	}
      }
    }
  }
}
\else
\newcommand\createtodoauthor[2]{%
\expandafter\newcommand\csname #1\endcsname[2][]{##1}%
}%
\fi

\definecolor{pairedNegOneLightGray}{HTML}{cacaca}
\definecolor{pairedNegTwoDarkGray}{HTML}{827b7b}
\definecolor{pairedOneLightBlue}{HTML}{a6cee3}
\definecolor{pairedTwoDarkBlue}{HTML}{1f78b4}
\definecolor{pairedThreeLightGreen}{HTML}{b2df8a}
\definecolor{materialYellow}{HTML}{ffe082}
\definecolor{materialDarkGreen}{HTML}{009d28}
\definecolor{materialLightGreen}{HTML}{daefc6}
\definecolor{materialLightGray}{HTML}{dfdfdf}
\definecolor{materialLightGrayBlueColor}{rgb}{0.95, 0.95, 0.97}
\definecolor{pairedFourDarkGreen}{HTML}{33a02c}
\definecolor{pairedFiveLightRed}{HTML}{fb9a99}
\definecolor{pairedSixDarkRed}{HTML}{e31a1c}

\createtodoauthor{grosser}{pairedOneLightBlue}
\createtodoauthor{sid}{pairedTwoDarkBlue}
\createtodoauthor{ag}{pairedThreeLightGreen}
\createtodoauthor{seb}{pairedFourDarkGreen}
\createtodoauthor{chris}{pairedFiveLightRed}
\createtodoauthor{authorSix}{pairedSixDarkRed}
\createtodoauthor{alex}{pairedNegTwoDarkGray}

\definecolor{PinkColor}{rgb}{1,0.2,0.6}
\definecolor{AlmostWhiteColor}{rgb}{0.9, 0.9, 0.9}
\definecolor{LightGrayColor}{rgb}{0.95, 0.95, 0.97}
\definecolor{MediumGrayColor}{rgb}{0.85, 0.85, 0.87}
\definecolor{DarkGrayColor}{rgb}{0.75, 0.75, 0.77}

\graphicspath{{./images/}}

\makeatletter
\newcommand\requiredelimiter[2][########]{%
	\ifdefined#2%
		\def\@temp{\def#2#1}%
		\expandafter\@temp\expandafter{#2}%
	\else
		\@latex@error{\noexpand#2undefined}\@ehc
	\fi
}
\@onlypreamble\requiredelimiter
\makeatother

\newcommand\newdelimitedcommand[2]{
	\expandafter\newcommand\csname #1\endcsname{#2}
	\expandafter\requiredelimiter
	\csname #1 \endcsname
}

\newcommand{\messa}[1]{\ensuremath{\texttt{LeanMLIR}(#1)}}

\newcommand{\stxlet}{\ensuremath{\texttt{let}}}
\requiredelimiter{\stxlet}

\newdelimitedcommand{toolname}{Tool}

\usepackage{booktabs}

\newcommand{\blind}[1]{XXXX} 

\newcommand{\mathlib}{\texttt{mathlib}\xspace}

\newcommand{\GreenBox}[1]{\smash{\colorbox{materialLightGreen}{\texttt{#1}}}}

\newcommand{\LightGrayBlueBox}[1]{\smash{\colorbox{materialLightGrayBlueColor}{\texttt{#1}}}}

\usepackage{tikz}
\usetikzlibrary{arrows}
\usetikzlibrary{shapes}
\DeclareRobustCommand\circled[2][]{\ifthenelse{\isempty{#1}}{\tikz[baseline=(char.base)]{\node[shape=circle,fill=pairedOneLightBlue,inner sep=1pt] (char) {#2};}}{\autoref{#1}: \hyperref[#1]{\tikz[baseline=(char.base)]{\node[shape=circle,fill=pairedOneLightBlue,inner sep=1pt] (char) {#2};}}}}

\usepackage[verbose]{newunicodechar}
\newunicodechar{Γ}{\ensuremath{\Gamma}}
\newunicodechar{⊢}{\ensuremath{\vdash}}
\newunicodechar{▸}{\ensuremath{\blacktriangleright}}
\newunicodechar{∅}{\ensuremath{\emptyset}}
\newunicodechar{α}{\ensuremath{\alpha}}
\newunicodechar{β}{\ensuremath{\beta}}
\newunicodechar{δ}{\ensuremath{\delta}}
\newunicodechar{Δ}{\ensuremath{\Delta}}
\newunicodechar{ϵ}{\ensuremath{\epsilon}}
\newunicodechar{τ}{\ensuremath{\tau}}
\newunicodechar{ε}{\ensuremath{\epsilon}}
\newunicodechar{σ}{\ensuremath{\sigma}}
\newunicodechar{Σ}{\ensuremath{\Sigma}}
\newunicodechar{∈}{\ensuremath{\in}}
\newunicodechar{∧}{\ensuremath{\land}}

\newunicodechar{₀}{\textsubscript{0}}
\newunicodechar{₁}{\textsubscript{1}}
\newunicodechar{₂}{\textsubscript{2}}
\newunicodechar{₃}{\textsubscript{3}}
\newunicodechar{ₙ}{\textsubscript{n}}
\newunicodechar{ₘ}{\textsubscript{m}}
\newunicodechar{ₕ}{\textsubscript{h}}
\newunicodechar{⊕}{\ensuremath{\oplus}}
\newunicodechar{∀}{\ensuremath{\forall}}
\newunicodechar{∃}{\ensuremath{\exists}}
\newunicodechar{∘}{\ensuremath{\circ}}
\newunicodechar{⟦}{\ensuremath{\llbracket}}
\newunicodechar{⟧}{\ensuremath{\rrbracket}}
\newunicodechar{ℕ}{\ensuremath{\mathbb{N}}}
\newunicodechar{ℤ}{\ensuremath{\mathbb{Z}}}

\newunicodechar{⦃}{\ensuremath{\{\{}}
\newunicodechar{⦄}{\ensuremath{\}\}}}
\newunicodechar{⧸}{\ensuremath{/}}
\newunicodechar{⊑}{\ensuremath{\sqsubseteq}}

\def\grantsponsor#1#2#3{#2}
\newcommand\grantnum[3][]{#3%
  \def\@tempa{#1}\ifx\@tempa\@empty\else\fi}

\newcommand{\fitree}{\textsf{Fitree}}
\requiredelimiter{\fitree}
\newcommand{\nt}{\leadsto}
\requiredelimiter{\nt}

\usepackage{pifont}
\newcommand{\numalivepatches}{700}

\author{Siddharth Bhat}{Cambridge University, United Kingdom}{sb2743@cam.ac.uk}{https://orcid.org/0009-0007-6410-3681}{}
\author{Alex Keizer}{Cambridge University, United Kingdom}{ack55@cam.ac.uk}{https://orcid.org/0000-0002-8826-9607}{}
\author{Chris Hughes}{University of Edinburgh, United Kingdom}{chughes6@ed.ac.uk}{}{}
\author{Andr\'{e}s Goens}{University of Amsterdam, Netherlands}{a.goens@uva.nl}{https://orcid.org/0000-0002-0409-1363}{}
\author{Tobias Grosser}{Cambridge University, United Kingdom}{tobias.grosser@cst.cam.ac.uk}{https://orcid.org/0000-0003-3874-6003}{}

\authorrunning{S. Bhat, A. Keizer, C. Hughes, A. Goens and T. Grosser} 

\Copyright{Siddharth Bhat, Alex Keizer, Chris Hughes, Andr\'{e}s Goens and Tobias Grosser} 



\category{} 

\relatedversion{}

\supplement{}
\supplementdetails[subcategory={}, cite={}, swhid={}]{Software}{https://zenodo.org/records/11519739}

\acknowledgements{We thank Sébastien Michelland and Sebastian Ullrich for their early help in this project and feedback, as well as Anton Lorenzen for his helpful feedback.}

\funding{ This project has received funding from the
\grantsponsor{horizonEU}{European Union's Horizon EUROPE research and innovation program}{https://www.horizon-eu.eu/} under grant agreement no. \grantnum{horizonEU}{101070374 (CONVOLVE)}.}

\EventEditors{John Q. Open and Joan R. Access}
\EventNoEds{2}
\EventLongTitle{The International Conference on Interactive Theorem Proving (ITP 2024)}
\EventShortTitle{ITP 2024}
\EventAcronym{ITP}
\EventYear{2024}
\EventDate{September 9--24, 2024}
\EventLocation{Tbilisi, Georgia}
\EventLogo{}
\SeriesVolume{42}
\ArticleNo{23}

\newenvironment{ottdefnblock}[3][]{ \framebox{\mbox{#2}} \quad #3 \\[0pt]}{}

\newcommand{\ottafterlastrule}{\\}

\title{Verifying Peephole Rewriting In SSA Compiler IRs}
\titlerunning{Verifying Peephole Rewriting In SSA Compiler IRs}

\nolinenumbers 

\begin{document}
\maketitle

\begin{abstract}
There is an increasing need for domain-specific reasoning in modern compilers.
This has fueled the use of tailored intermediate representations (IRs) based on static single assignment (SSA), like in the MLIR compiler framework.
Interactive theorem provers (ITPs) provide strong guarantees for the end-to-end verification of compilers (e.g., CompCert).
However, modern compilers and their IRs evolve at a rate that makes proof engineering alongside them prohibitively expensive.
Nevertheless, well-scoped push-button automated verification tools such as the Alive peephole verifier for LLVM-IR gained recognition in domains where SMT solvers offer efficient (semi) decision procedures.
In this paper, we aim to combine the convenience of automation with the versatility of ITPs for verifying peephole rewrites across domain-specific IRs.
We formalize a core calculus for SSA-based IRs that is generic over the IR and covers so-called regions (nested scoping used by many domain-specific IRs in the MLIR ecosystem).
Our mechanization in the Lean proof assistant provides a user-friendly frontend for translating MLIR syntax into our calculus.
We provide scaffolding for defining and verifying peephole rewrites, offering tactics to eliminate the abstraction overhead of our SSA calculus.
We prove correctness theorems about peephole rewriting, as well as two classical program transformations.
To evaluate our framework, we consider three use cases from the MLIR ecosystem that cover different levels of abstractions:
(1) bitvector rewrites from LLVM, (2) structured control flow, and (3) fully homomorphic encryption.
We envision that our mechanization provides a foundation for formally verified rewrites on new domain-specific IRs.

\end{abstract}

\section{Introduction}
\label{sec:intro}

Static single assignment (SSA)~\cite{rastello2022ssa} intermediate representations (IRs) are at the core of modern compilers, thanks to the benefits their immediate encoding of use-def relationships brings to compiler analyses and transformations.
\emph{Peephole optimizations}~\cite{mckeeman1965peephole}, which replace assembly-level instruction sequences of bounded length with semantically equivalent optimized ones, benefit from SSA during target code generation~\cite{10.1145/151333.151368} and are now also widely used for optimizing SSA-based IRs.
Peephole optimizations are so common, that 10\% of all IR transforming code in LLVM~\cite{lattner2004llvm} belongs to its InstCombine peephole optimizer,\footnote{Non-blank and non-comment lines of \texttt{.cpp} files in \texttt{llvm/lib/Transforms} on commit \texttt{f4f1cf6c3}.} which is beyond the size of LLVM's loop optimizer.
Further evidence is offered by LLVM's commit log, where the most referenced tool is the Alive peephole verifier~\cite{lopes2021alive2}. Alive has brought automatic SMT-based verification into the day-to-day of the LLVM compiler community.

In the context of the end-to-end verified compiler Compcert~\cite{leroy2016compcert}, peephole rewriting has been formalized (and mechanized) in its classical form of straight-line assembly code~\cite{mullen2016verified}, but this verification does not cover rewriting along the SSA def-use chain.
As an example, consider the rewrite $(y = x + 1; z = y - 1) \mapsto (z = x)$.
This pattern does \emph{not} match the program $(y = x + 1; \textbf{p = y}; z = y - 1)$ in straight-line rewriting, due to the interleaved instruction $p = y$.
On the other hand, by concentrating on the dataflow, we rewrite any subprogram of the form $(y = x + 1;~\Circle~; z = y - 1)$ to $(y = x + 1;~\Circle~; z = x)$, regardless of what fills the hole $\Circle$.
This rewriting on the ``def-use'' chain can be applied to assembly code before register allocation and all SSA-based IRs.

SSA-based IRs have been successful in domain-specific compilers, where they enable concise reasoning at the favored abstraction level.
In particular, the MLIR compiler framework~\cite{lattner2020mlir} has been widely adopted for machine learning~\cite{linalg}, quantum computing~\cite{peduri2022qssa}, and even for compiling Lean~\cite{bhat2022lambda}. MLIR lowers the cost of instantiating domain-specific IRs and encourages transformations on specialized high-level IRs. 
Instead of complex potentially side-effectful global reasoning at a lower abstraction level, these tailored IR abstractions often offer value semantics (i.e., referential transparency) to enable side-effect-free local reasoning.
MLIR also introduces the concept of regions, which allow IR operations to be nested, enabling structured control flow. Structured control makes termination proofs of loops easier and the tailored domain-specific IRs have the potential to reduce the complexity of proofs.

In this paper, we verify peephole rewriting over SSA-based IRs.
We formalize a core calculus for SSA-based IRs that is generic over the IR and covers regions instead of potentially diverging unstructured control. We mechanize our calculus in the Lean~\cite{demoura2021lean} proof assistant and make it accessible to MLIR developers by offering an embedding of MLIR syntax.
Concretely, our contributions are:
\begin{itemize}
  \item A formalization of SSA with regions parametrized over a user-defined IR $X$ and its mechanization in our framework\footnote{Our code is available at \url{https://github.com/opencompl/lean-mlir/releases/tag/ITP24} (ae0dd933).} \messa{X} that exploits denotational-style value semantics for optimizing along the SSA use-def chain of an MLIR-style IR (Sections~\ref{sec:mlir-intro}, \ref{sec:mlirnatural-under-the-hood})
  \item Evidence that our formalization of SSA allows for effective meta-theoretic reasoning:
  \begin{itemize}
	\item A verified peephole rewriter, for which we prove that lifting a peephole rewrite to a rewrite on the entire program preserves semantics (\autoref{sec:verified-rewriting})
    \item Two verified implementations of generic SSA-based optimizations: dead code elimination and common subexpression elimination (\autoref{sec:dce})
    \item Proof automation for eliminating the abstraction overhead of our SSA calculus and exposing clean mathematical proof obligations for each rewrite (\autoref{sec:automation})
\end{itemize}
   \item An extension of our pure optimizations in a context with side effects (\autoref{sec:sideeffects})
   \item Syntax, semantics, and local rewrites for three MLIR-based IRs:
	      (1) arithmetic over bitvectors, (2) structured control flow, and (3) fully homomorphic encryption (\autoref{sec:casestudies})
\end{itemize}

\begin{figure}[t]
\begin{subfigure}[b]{\textwidth}
\begin{minipage}[t]{0.35\textwidth}
\begin{lean}
inductive Ty
| r
| nat
\end{lean}
\end{minipage}
\begin{minipage}[t]{0.64\textwidth}
\begin{lean}
inductive Op
| arith_const (x : Nat) -- with compile-time data `x`
| monomial -- build equivalence class of monomial
| add -- add op.
\end{lean}
\end{minipage}
\caption{\label{fig:quotring-op-ty}
\texttt{QuotRing} has three
\texttt{Op} constructors, \texttt{add}, \texttt{monomial}, and \texttt{arith\_const x} (for \texttt{x} an element of $\mathbb{N}$) matching the three operations of the IR and two 
\texttt{Ty} constructors, \texttt{r} and \texttt{nat} matching the two IR types.}
	\end{subfigure}\\

	\begin{subfigure}[b]{\textwidth}
		\begin{lean}

instance : OpSignature Op Ty where signature
  | .arith_const _ => { sig := [], outTy := .nat } -- takes no args, returns `nat`
  | .add      => { sig := [.r, .r], outTy := .r } -- takes two `r`, returns `r`
  | .monomial => { sig := [.nat, .nat], outTy := .r } -- takes two`.nat`, returns `r`
		\end{lean}
		\caption{\label{fig:quotring-syntax} User-defined signatures of each \texttt{QuotRing} operation.}
	\end{subfigure}\\

	\begin{subfigure}[b]{\textwidth}
\begin{lean}
noncomputable def generator : (ZMod q)[X] := X^(2^n) + 1
abbrev R := (ZMod q)[X] ⧸ (span {generator q n})

instance : TyDenote Ty where
  toType
  | .r => R -- the denotation of `r` is an element of the ring `R`
  | .nat => Nat

instance : OpDenote Op Ty where
  denote
  | .arith_const (x : Nat), _, _ => x -- the denotation of `arith_const x` is  `x`
  | .add, [(x : R), (y : R)]ₕ, _ => x + y
  | .monomial, [(c : Nat), (i : Nat)]ₕ, _ =>
        Quotient.mk (span {generator q n}) (monomial i c)
\end{lean}
		\caption{\label{fig:quotring-semantics} User-defined semantics of \texttt{QuotRing}.
			The \texttt{instance} syntax is used to define a typeclass instance, by specifying the corresponding members, which in this case are the denotation functions.
			The \texttt{noncomputable} annotation in Lean tells the compiler not to generate executable code for this function since \mathlib uses a noncomputable definition for quotients of polynomial rings.
			Note that our framework ensures that values are well-typed according to \texttt{OpSignature} and \texttt{TyDenote}.}
	\end{subfigure}
	\caption{\label{fig:quotring-defs}User definitions for \texttt{QuotRing}, which declares the operations and types of the IR, the type signatures of the operations, and the denotations of the types and operations into Lean types.}
\end{figure}

\section{Motivation: Verfying Optimizations for High-Level IRs}
\label{sec:mlir-intro}

Effective domain-specific optimizations are almost impossible when reasoning on traditional LLVM-style compiler IRs.
These offer a ``universal'' low-level abstraction, originally designed to represent C-style imperative code.
Such LLVM-style IRs are built around the concepts of load/store/arithmetic/branching, which is ideal when optimizing at the level of scalar arithmetic, instruction scheduling, or applying certain kinds of loop optimizations.
However, this level of abstraction is unsuitable for reasoning about high-level mathematical abstractions.

Consider a compiler for Fully Homomorphic Encryption (FHE)~\cite{gentry2009fully}, a cryptographic technique that uses algebraic structures to allow an untrusted third party to do computation on encrypted data.
In such a compiler, we might have a rewrite like $(a + X^{2^n} + 1 \mapsto a)$, which is a simple identity on the corresponding quotient ring.\footnote{We will discuss the underlying mathematical structure in more detail in~\autoref{sec:fully-homomorphic-encryption}}
Expressed in LLVM, the computation of this simple operation consists of multiple basic blocks forming a loop, each containing memory loads, pointer arithmetic, scalar operations, and branches.
As a result, the algebraic structure is completely lost and exploiting simple algebraic identities turns into a heroic effort of reasoning about side effects and stateful program behavior.
State-of-the-art compilers for FHE consequently use domain-specific IRs (often expressed with MLIR~\cite{viand2022heco,heaanmlir}) when generating optimized code for FHE, where algebraic optimizations can take place at an FHE-specific IR that has value-semantics (e.g., is referentially transparent) and is overall closer to the mathematical structure of the problem.

\subsection{Defining \messa{\operatorname{QuotRing}}: Syntax and Semantics}

As an example, we model an IR aimed at FHE that manipulates objects in the algebraic structure $R \equiv (\mathbb{Z}/q\mathbb{Z})[X] / (X^{2^n} + 1)$.
To model it, we instantiate an IR \messa{\operatorname{QuotRing}} in our framework.
It has three simple operations: \texttt{arith\_const} and \texttt{monomial}, to construct values in $R$, and \texttt{add} to add two values of $R$.
To define the syntax and semantics of \messa{\operatorname{QuotRing}}, we first declare the types and operations in the IR (\autoref{fig:quotring-op-ty}).
\texttt{QuotRing} has two types: \texttt{r}, which represents the ring $R$, and $\texttt{nat}$ for naturals.
Terms in \texttt{Op} represent the operations \texttt{arith\_const}, \texttt{monomial}, and \texttt{add}, as well as associated compile-time data.
We then define the operation signatures by giving an instance of the \texttt{OpSignature} typeclass, which is offered by our framework to instantiate custom IRs (\autoref{fig:quotring-syntax}).
That is, for each operation we specify:
(1) the arity and types of arguments \texttt{sig}, and
(2) the type of the return value \texttt{outTy}.
The operation \texttt{arith\_const} takes no arguments and returns a \texttt{nat}, \texttt{monomial} and \texttt{add} take two \texttt{nat}/\texttt{r}-valued arguments respectively, and both return an \texttt{r}.

The type denotation is also simple to express with the \texttt{TyDenote} typeclass (\autoref{fig:quotring-semantics}).
\texttt{Ty} thus represents the embedded type in the IR and has only two inhabitants \texttt{r} and \texttt{nat}, whose denotation are \texttt{R} and \texttt{Nat}, the Lean (host) type that represents the mathematical objects $R$ and $\mathbb{N}$ respectively.
The denotation of operations is a Lean function from the denotation of the input types (as recorded in the signature of that operation), to the denotation of the output type.\footnote{Our framework groups type and operation denotations into a \texttt{Dialect}, which we leave out for brevity.}
Concretely, an \texttt{arith\_const n} operation takes no arguments, so its denotation is Lean's \texttt{Nat}, while \texttt{add} takes two \texttt{r} arguments, so its denotation is a function from the product\footnote{The mechanization uses a heterogeneous vector type \texttt{HVector}, which is coerced into the product type.} of its arguments to its output, i.e., \texttt{R × R → R}.
The same is true for \texttt{monomial} for \texttt{Nat × Nat → R}.
We define the denotation of  \texttt{arith\_const n} to evaluate to \texttt{n}, \texttt{add(x, y)} to evaluate to \texttt{x~+~y}, and \texttt{monomial(a, i)} to evaluate to \texttt{Quotient.mk (span {generator p q} (monomial a i))}, the equivalence class of the monomial $a X^i$.
As the \texttt{QuotRing} IR does not require regions (\autoref{sec:structured-control-flow}), we only need to add MLIR syntax support (\autoref{sec:writing-mlirnatural-programs-in-mlir-syntax}) and translate the MLIR AST to \texttt{Ty} and \texttt{Op}, e.g., mapping \texttt{!Nat}\footnote{In practice, one would use a fixed-bitwidth type \texttt{iN} but we use \texttt{!Nat} for a simpler exposition.} to \texttt{nat} and \texttt{!R} to \texttt{r}, and a full \texttt{QuotRing} example (\autoref{fig:quotring-peephole-rewrite-example}) can be written in Lean.

\subsection{Defining and Executing Peephole Rewrites for \texttt{QuotRing}}
\label{sec:rewriter}

We now verify the peephole rewrite $(a + X^{2^n} + 1 \mapsto a)$, where a is a variable and $X^{2^n}$ is a constant in the ring.
In $(\mathbb{Z}/q\mathbb{Z})[X] / (X^{2^n} + 1)$ this rewrite is simple to prove and, unsurprisingly, our custom \messa {{\operatorname{QuotRing}}} IR enables us to rewrite at exactly this level.
Any given peephole rewrite (of which \autoref{fig:quotring-peephole-rewrite-example} is an example) consists of a context $\Gamma$ of free variables in the search pattern of the peephole rewrite.
The search pattern is called \texttt{lhs}, and the replacement is \texttt{rhs}.
The user has a proof obligation that the denotations of the left and right-hand sides are equal, which is given by the field \texttt{correct} of the peephole rewrite.
In later examples, we reason upto semantic refinement to incorporate LLVM's notion of poison values \cite{lee2017taming}.
For now, we stick to equality to simplify exposition.

We declare our desired peephole rewrite by defining \texttt{a\_plus\_generator\_eq\_a}.
Its type is \texttt{PeepholeRewrite Op [.r] r}, where the \texttt{Op} specifies the IR the rewrite belongs to and \texttt{[.r]} is the list of types of free variables in the program.
For $(a + X^{2^n} + 1 \mapsto a)$, this is ($a : \texttt{r}$).
The final instruction we are matching yields a value of type \texttt{r}.
The \texttt{lhs} is the program fragment we want to match on, with the free variable \texttt{\%a} interpreted as being allowed to match any variable of type \texttt{r}.
Observe that the type encapsulates exactly what is necessary for a well-typed match:
the types of free variables \texttt{r} and the type of the instruction whose return value we are replacing (also \texttt{r} in this case).
The rewritten program is the \texttt{rhs} field.
\begin{figure}
	\begin{lean}
def a_plus_generator_eq_a : PeepholeRewrite Op [.r] .r := {
  lhs /- a + X^(2^n) + 1 -/ := [quotring_com q, n| {
	^bb0(%a : !R):
	  %one_int = arith.const 1 : !Nat
	  %two_to_the_n = arith.const ${2**n} : !Nat
	  %x2n = poly.monomial %one_int, %two_to_the_n : (!Nat, !Nat) -> !R
	  %oner = poly.const 1 : !R
	  %p = poly.add %x2n, %oner : !R
	  %v1 = poly.add %a, %p : !R
	  return %v1 : !R
  }],
  rhs /- a -/ := [quotring_com q, n| {
    ^bb0(%a : !R):
      return %a : !R
  }],
  correct := by
    funext Γv; simp_peephole [Nat.cast_one, Int.cast_one] at Γv @\circled{1}@
    /- ⊢ a + ((Quotient.mk (span {f q n})) ((monomial (2**n)) 1) + 1) = a -/
    ... /- simple proof with only definitions and theorems from Mathlib -/
}
	\end{lean}
	\caption{\label{fig:quotring-peephole-rewrite-example}
	A peephole rewrite in \messa{\operatorname{QuotRing}} asserts the semantic equivalence of two SSA programs given in MLIR syntax.
	Our proof automation through \texttt{simp\_peephole} eliminates the framework overhead, such that closing a clean mathematical goal suffices to prove correctness.
	}
\end{figure}

Both the left- and right-hand sides of the rewrite are written in MLIR syntax.
Note that we also include a custom quasiquotation \texttt{\$\{2**n\}}, to inline the symbolic (universally quantified) value $n$, even though the IR would require $2^n$ to be a concrete constant.
Using MLIR syntax matches the LLVM community's use of automation tooling, such as Alive: copy a code snippet and get a response.
Our goal is to make the use of an interactive theorem prover part of the day-to-day workflow of compiler engineers.
To enable this workflow, we implement a full MLIR syntax parser, along with facilities to convert from the generic MLIR abstract syntax into our framework type, such that we can use MLIR syntax in Lean.

To prove the correctness of \texttt{a_plus_generator_eq_a}, we use the \texttt{simp_peephole} \circled{1} tactic from our framework, which removes any overhead of our SSA encoding. We are left with:
\LightGrayBlueBox{\texttt{⊢ a + ((Quotient.mk (span {f q n})) ((monomial (2**n)) 1) + 1) = a}},
a proof obligation in the underlying algebraic structure that, thanks to Lean's \mathlib, can be closed with a few (elided) lines of algebraic reasoning.

\subsection{Executing Peephole Rewrites}

Given a peephole rewrite \texttt{rw} and a source program \texttt{s}, we provide \texttt{rewritePeephole} to replace the pattern \texttt{rw.lhs} in the source program \texttt{s}.
If the matching succeeds, we insert the target program \texttt{rw.rhs} (with appropriate substitutions) and replace all references to the original variable with a reference to the newly inserted \texttt{var}.
Note that the matching is based on the def-use chain.
Thus, a pattern need not be \emph{syntactically} sequential in the program \texttt{s}.
As long as the pattern \texttt{rw.lhs} can be found as a \emph{subprogram} of \texttt{s}, \texttt{s} will be rewritten.
This makes our peephole rewriter an SSA peephole rewriter, which distinguishes it from a straight-line peephole rewriter that only matches a linear sequence of instructions.

Thanks to our intrinsically well-typed encoding, we know that the result of the rewriter is always a well-typed program, under the same context and resulting in the same type as the original program.
Furthermore, the framework extends the local proof of semantic equivalence to a global proof, showing that the rewriter is semantics preserving:

\begin{lean}
/- The denotation of the rewritten program is equal to the source program. -/
theorem denote_rewritePeephole (fuel : ℕ) (rw : PeepholeRewrite Op Γ t)
  (target : Com Op Γ₂ t₂) : (rewritePeephole fuel rw target).denote = target.denote
\end{lean}

These typeclass definitions are all we need to define the \texttt{QuotRing} IR.
Our framework takes care of building the intrinsically well-typed IR for \texttt{QuotRing} from this, and gives us a verified peephole rewriter, with other optimizations like CSE and DCE.
We will now delve into the details of the framework and see how it achieves this.

\section{\messa{X}: A Framework for Intrinsically Well-Typed SSA}
\label{sec:mlirnatural-under-the-hood}

In this section, we describe the core design of the framework: the encoding of programs and their semantics in \messa{X} (\autoref{fig:iexpr-icom-def}).
We review some dependently-typed tooling we use to define our IR.
\textbf{Contexts:}
Our encoding is intrinsically well-typed (i.e., each inhabitant of \texttt{Expr} or \texttt{Com} described below is, by construction, well typed).
Thus, we need a \emph{context} to track the types of variables that are allowed to occur (\texttt{Ctxt Ty}).
A context is a list of types, where for example \texttt{[int, int, bool]} means that there are two variables of the (user-defined) type \texttt{int} and one variable of type \texttt{bool} we may refer to.
\textbf{Variables:}
The type \texttt{(Var Γ α)} encodes variables of type \texttt{α} in context \texttt{Γ}.
We use De Bruijn indices~\cite{pierce2002types} in the standard way, but, additionally, a variable with index $i$ also carries a proof witness that the $i$-th entry of context \texttt{Γ} is the type \texttt{α}.
\textbf{Heterogeneous Vectors:} To define an argument signature (\texttt{OpSignature.sig}), say, \texttt{[int, int, bool]}, we need an expression with this operation to store two variables of type \texttt{int} and one of type \texttt{bool}.
We want to statically ensure that the types of these variables are correct, so we store them in a heterogeneous vector. A vector of type \texttt{HVector f [α₁, ..., αₙ]}
is equivalent to a tuple \texttt{(f α₁ × ... × f αₙ)}.

\subsection{Semantics of \messa{X}}
The core types for programs are \texttt{Expr} and \texttt{Com}, shown in \autoref{fig:iexpr-icom-def}.
The type \texttt{Expr~Γ~α} describes individual SSA operations; we think of it as a function from values in the context \texttt{Γ}---also called a \emph{valuation} for that context---to a value in the denotation of type \texttt{α}.
The type \texttt{Com~Γ~α} has a similar interpretation but represents sequences of operations.
Each command binds a new value in the current context (the \texttt{var} constructor) until the sequence returns the value of one such variable \texttt{v} (the \texttt{ret} constructor).
Thus, this encoding of SSA exploits the similarity to the ANF~\cite{appel1998ssa}
and CPS~\cite{kelsey1995correspondence} encodings.
In particular, our \texttt{Expr} represents an SSA assignment, and \texttt{Com}s represents a block of operations, often called a basic block. A basic block typically would either return or branch to another block. In our case, blocks only return and consequently do not model branching. Instead, we use regions to model structured control flow (\autoref{sec:structured-control-flow}).
Given our core syntax, our framework now automatically expands the semantics given by the user in \GreenBox{\texttt{OpDenote}} to semantics for \texttt{Expr} and \texttt{Com} (\autoref{fig:iexpr-icom-denote}).
An \texttt{Expr} evaluates its arguments by looking up their value in the valuation and then invokes the user-defined \texttt{OpDenote.denote} to evaluate the semantics of the \texttt{op}.

\begin{figure}
\begin{subfigure}[b]{\textwidth}
\begin{lean}

inductive Expr [OpSignature Op Ty] : Ctxt Ty → Ty → Type where
| mk (op : Op) -- op (arg1, arg2, ..., argn) : outTy op
  (args : HVector (Var Γ) (OpSignature.sig op)) : Expr Γ (OpSignature.outTy op)

inductive Com [OpSignature Op Ty] : Ctxt Ty → Ty → Type where
| ret (v : Var Γ α) : Com Γ α -- return v
| var (e : Expr Γ α) (body : Com (Γ.snoc α) β) : Com Γ β -- let v : α := e in body
\end{lean}
\vspace{-1em}
\caption{\label{fig:iexpr-icom-def} Core syntax of \messa{X}, polymorphic over \texttt{Op}.
The arguments in square brackets are assumed typeclass instances. \texttt{Type} is the base universe of Lean types. }
\end{subfigure}\\
\vspace{2em}
\begin{subfigure}[b]{\textwidth}
\begin{lean}
variable [TyDenote Ty] [OpDenote Op Ty] [DecidableEq Ty]

def Expr.denote : {ty : Ty} → (e : Expr Op Γ ty) → (Γv : Valuation Γ) → toType ty
| _, ⟨op, args⟩, Γv => @\GreenBox{OpDenote.denote}@ op (args.map (fun _ v => Γv v))

def Com.denote : Com Op Γ ty → (Γv : Valuation Γ) → (toType ty)
| .ret e, Γv => Γv e
| .var e body, Γv => @\GreenBox{body.denote}@ (Γv.snoc (@\GreenBox{e.denote}@ Γv))
\end{lean}
\vspace{-1em}
\caption{\label{fig:iexpr-icom-denote}Denotation of \texttt{Expr} and \texttt{Com} in \messa{X}, which extends the user's \texttt{OpDenote}
to entire programs. Intrinsic well-typing of \texttt{Com} makes its denotation
a well-typed function from the context valuation to the return type. The angled brackets are used to pattern match on a structure constructor anonymously. }
\end{subfigure}
\vspace{-3em}
\caption{Definitions in \messa{X} for \texttt{Expr} and \texttt{Com}, and their associated denotations.}
\end{figure}

\subsection{Writing \messa{X} Programs Using MLIR Syntax}
\label{sec:writing-mlirnatural-programs-in-mlir-syntax}

An important goal for our framework is to provide easy access to formalization for the MLIR community.
Toward this goal, we have a deep embedding of MLIR's AST and a corresponding parser.
This is developed using Lean's syntax extensions~\cite{ullrich2020beyond}.
We extend Lean with a generic framework to build \texttt{Expr} and \texttt{Com} terms from a raw MLIR AST.
This framework allows the user to pattern-match on the MLIR AST to build intrinsically well-typed terms, as well as to throw errors on syntactically correct, but malformed MLIR input.
These are used by our framework to automatically convert MLIR syntax into our SSA encoding, along with the ability to provide precise error messages in cases of translation failure.
This enables us to write all our examples in MLIR syntax, as demonstrated throughout the paper.

\newcommand{\MLIRAstLinesOfCode}{around 400 lines of code}
\newcommand{\MLIREDSLLinesOfCode}{around 900 lines of code}
\newcommand{\MLIRTranslationFrameworkLinesOfCode}{around 450 lines of code}
\newcommand{\MLIRTranslationSimpleIRLinesOfCode}{around 50 lines of code}

More concretely, we have an embedded domain-specific language (EDSL), which declares the MLIR grammar as a Lean syntax extension.
As part of this work, we have found several inconsistencies with the MLIR language reference and contributed patches upstream to update them.\footnote{reviews.llvm.org/\{D122979, D122978, D122977, D119950, D117668\}}
Overall, this gives users the ability to write idiomatic MLIR code into our framework and receive an MLIR AST.
Moreover, as we will showcase in the examples, our EDSL is idiomatically embedded into Lean, which allows us to quasiquote Lean terms.
This will come in handy to write programs that are generic over constants, such as parameterizing a program by $2^n$ for any choice of $n$.
We build our intrinsically well-typed data structures from this MLIR AST by writing custom elaborators.
\subsection{Modelling Control Flow in \messa{X} With Regions}
\label{sec:modelling-control-flow-in-mlirnatural}

So far, our definition of \texttt{Com} only allows straight-line programs.
To be able to model control flow, we add regions to our IR.
Regions are an extension to SSA introduced by MLIR. They add the syntactic ability to nest IR definitions, thereby allowing syntactic encoding of concepts such as structured control flow.
This is in contrast with the approach of having a sea of basic blocks in a control-flow graph (CFG) that are connected by branch instructions.
More specifically, structured control flow with regions allows modeling reducible control flow~\cite{aho1985compilers}.
General CFGs allow us to represent more complex, irreducible control flow, which makes them harder to reason about.
Consequently, compiler frameworks such as MLIR encourage structured control flow (even though they allow for a sea of basic blocks). In our framework, we focus on the novel aspects of MLIR: structured control via nested regions.

\begin{figure}
\begin{leanfootnotesize}
structure OpSignature (Ty : Type) where /- (1) Extending signature. -/
@\GreenBox{regSig : List (Ctxt Ty × Ty)}@
@$\cdots$@

class OpDenote [TyDenote Ty] [OpSignature Op Ty] where /- (2) Extending denotation. -/
denote : (op : Op) → (args : HVector toType (OpSignature.sig op)) →
@\GreenBox{(regArgs : HVector (fun (ctx, t) => Valuation ctx → toType t) (OpSignature.regSig op))}@ →
(toType (OpSignature.outTy op))

inductive Expr : (Γ : Ctxt Ty) → (ty : Ty) → Type where
| mk (op : Op)
@$\cdots$@
@\GreenBox{(regArgs : HVector (fun (ctx, ty) => Com ctx ty) (OpSignature.regSig op))}@:
Expr Γ ty

mutual /- (3) extending expression denotation to recursively invoke regions. -/
def Expr.denote : {ty : Ty} → (e : Expr Op Γ ty) → (Γv : Γ.Valuation) → (toType ty)
| _, ⟨op, args, regArgs⟩, Γv =>
OpDenote.denote op (args.map (fun ty v => Γv v)) @\GreenBox{regArgs.denote}@
@$\cdots$@
end
\end{leanfootnotesize}
\vspace{-1em}
\caption{\label{fig:mechanization-regions} Extending \messa{X} with regions.
New fields are \GreenBox{in green}.
In \texttt{OpDenote}, one can now access the sub-computation represented by the region when defining the semantics of \texttt{Op}.}
\end{figure}

Intuitively, regions allow an \texttt{Op} to receive \texttt{Com}s as arguments, and choose to execute these \texttt{Com} arguments zero, one, or multiple times.
This allows us to model if conditions (by executing the regions zero or once), loops (by executing the region $n$ times), and complex operations such as tensor contractions and convolutions by executing the region on the elements of the tensor \cite{linalg}.
We implement this by extending \texttt{Expr} with a new field representing region arguments (\autoref{fig:mechanization-regions}).
We also extend \texttt{OpSignature} with an extra argument for the input types and output types of the region.
In parallel, we add the denotation of regions as an argument, extending \texttt{OpDenote}.
Similarly, we extend the denotation of \texttt{Expr} to compute the denotation of the region \texttt{Com}s in the \texttt{Expr}, before handing off to \texttt{OpDenote}.

This extension to our core calculus gives us the ability to model structured nesting of programs whose denotation is a bounded computation.\footnote{Since our semantics denote into Lean expressions, the user-given semantics must be provably terminating to be executable.
We wish to explore richer denotations, such as (computable) coinductive and (noncomputable) domain theoretic semantics in future work.}
This is used pervasively in MLIR, to represent \texttt{if} conditions, \texttt{for} loops, and higher-level looping patterns such as multidimensional strided array accesses over multidimensional arrays (tensors).
We show how to model control flow in \autoref{sec:structured-control-flow}.

\section{Reasoning About \messa{X}}
\label{sec:metatheory}

The correctness of peephole rewriting is a key aspect of the metatheory of \messa{X}.
We begin by sketching the mechanized proof of correctness of peephole rewriting.
We then discuss how the infrastructure built for this proof is reused to prove two other SSA optimizations:
common subexpression elimination (CSE) and dead code elimination (DCE).
Finally, we discuss our proof automation,
which manipulates the IR encoding at elaboration time to eliminate all references to the framework and provide a clean goal to the proof engineer.

\subsection{Verified SSA Rewriting With \texttt{rewritePeephole}}
\label{sec:verified-rewriting}

{\sloppypar
We now provide a sketch of the mechanized correctness proof of \texttt{rewritePeephole}.
The key idea is that to apply a rewrite at location $i$, we open up the \texttt{Com} at location $i$ in terms of a zipper~\cite{huet1997zipper}.
This zipping and rewriting at a location $i$ is implemented by \texttt{rewritePeepholeAt}.
The zipper comprises of \texttt{Lets} to the left-hand side of $i$, and \texttt{Com} to the right:}
\vspace{.5em}
{\small
\colorbox{MediumGrayColor}{
	\texttt{let $x_2$ = $x_1$; (let $x_3$ = $x_2$; (let $x_4$ = $x_3$; (return $x_3$))): Com $[x_1]$ α}
}
=

\quad \colorbox{DarkGrayColor}{
	\texttt{((let $x_2 = x_1$); let $x_3 = x_2$); : Lets $[x_1]$ $[x_1, x_2, x_3]$}
}

\quad \colorbox{pairedThreeLightGreen}{
\texttt{(let $x_4$ = $x_3$; (return $x_3$)) : Com $[x_1, x_2, x_3]$ α}
}
}

\vspace{.5em}

The use of a zipper enables us to easily traverse the sequence of let-bindings during transformation and exposes the current \texttt{let} binding being analyzed.
This exposing is performed by \texttt{Lets}, which unzips a \texttt{Com} such that the outermost binding of a \texttt{Lets} is the innermost binding of a \texttt{Com}.
This forms the zipper, which splices the \texttt{Com} into a \texttt{Com} and a \texttt{Lets}.
Also, while \texttt{Com} tracks only the return type $\alpha$ in the type index, \texttt{Lets} tracks the entire resulting context $\Delta$.
That is, in \texttt{(lets : Lets Γ Δ)}, the first context, \texttt{Γ}, lists all free variables (just as in \texttt{Com Γ t}), but the second context, \texttt{Δ}, consists of all variables in \texttt{Γ} plus a new variable for each let-binding in the sequence \texttt{lets}. We can thus think of \texttt{Δ} as the context at the current position of the zipper.
Another difference is the order in which these sequences grow. Recall that in \texttt{Com}, the outermost constructor represents the topmost let-binding. In \texttt{Lets}, the outermost constructor instead corresponds to the bottom-most let-binding.
This difference is what makes the zipper work.

{\sloppypar
	We have two functions to go from a program to a zipper and back:
	(1) \texttt{(splitProgramAt pos prog)}, to create a zipper from a program \texttt{prog} by moving the specified number of bindings to a new \texttt{Lets} sequence, and
	(2) \texttt{(addComInMiddleOfLetCom top mid bot)}, to turn a zipper \texttt{top, bot} into the program, while inserting a program \texttt{mid~:~Com} in between.
}
We also prove that the result of splitting a program with \texttt{splitProgramAt} is semantically equivalent to the original program.
Similarly, we prove that stitching a zipper back together with \texttt{addComInMiddleOfLetCom} results in a semantically equivalent program.

Given a peephole rewrite (\texttt{matchCom}, \texttt{rewriteCom}), to rewrite at location $i$, we first split the target program into \texttt{top} and \texttt{bot}.
We then attempt to match the def-use chain of the return variable in \texttt{matchCom} with the final variable in \texttt{top} (which is the target $i$, since we split the program there).
This matching of variables recursively matches the entire expression tree.\footnote{We match regions in expressions for structural equality. We \emph{do not} recurse into regions during matching, and treat regions as black-boxes.}
Upon successful matching, this returns a substitution $\sigma$ for the free variables in \texttt{matchCom} in terms of (free or bound) variables of \texttt{top}.
Using this successful matching, we stitch the program together as $\texttt{top}; \sigma(\texttt{rewriteCom}); \tau(\texttt{bot})$.
Here, $\tau$ is another substitution that replaces the variable at location $i$ with the return variable of $\texttt{rewriteCom}$.
Since we derived a successful matching, we know that the semantics of variable $i$ is equal to that of the return variable of \texttt{matchCom}.
By assumption on the peephole rewrite, the variable $i$ is equivalent to the return variable of \texttt{rewriteCom}.
This makes it safe to replace all occurrences of the variable $i$ in \texttt{bot} with the return variable of \texttt{rewriteCom}.
This proves \texttt{denote_rewritePeephole}, which states that if a rewrite succeeds, then the semantics of the program remain unchanged.
In this way, we use a zipper as a key inductive reasoning principle to mechanize the proof of correctness of SSA-based peephole rewriting. We extend this rewriting to regions by recursively rewriting over the regions in a program.

\subsection{DCE \& CSE: Folding Over Intrinsically Well Typed SSA}
\label{sec:dce}
\newcommand{\DeadCodeElimNumLinesCode}{$\approx 400$ LoC}
\newcommand{\DeadCodeElimNumLinesFramework}{$\approx 200$ LoC}
\newcommand{\CommonSubexprElimNumLinesCode}{$\approx 400$ LoC}

The classic optimizations enabled by SSA are peephole rewriting, dead code elimination (DCE), and common subexpression elimination (CSE).
We implement these optimizations in our framework as a test of its suitability for metatheoretic reasoning.
Our approach is different from previous approaches~\cite{zhao2013formal, barthe2014formal} with our use of intrinsic well-typing,
which mandates proofs of the structural rules on contexts to rewrite programs.
We begin by building machinery to witness that a context $Δ$ is equal to the context $\Gamma$, minus the variable $x$. This is spelled as $\texttt{Deleted}~\Gamma~x~\Delta$ in \messa{X}.
We then prove context-strengthening theorems to delete variables that do not occur in \texttt{Expr} and \texttt{Com} while preserving denotation.

Using this tooling, DCE is implemented in \DeadCodeElimNumLinesCode{}, which shows that our framework is well-suited to metatheoretic reasoning.
The implementation is written in a proof-carrying style, interleaving function definitions with their proof of correctness.
The recursive step of the dead code elimination takes a program $p : \texttt{Com Γ t}$ and a variable $v$ to be deleted, and returns a new $p': \texttt{Com Δ t}$.
The two contexts $\Gamma$ and $\Delta$ are linked by a context morphism \texttt{(Hom Γ Δ)}, to interpret $p'$ (with the deleted variable) which lives in a strengthened context $\Delta$ in the old context $\texttt{Γ}$.
We walk $p$ recursively to eliminate dead values at each \texttt{let} binding.
This produces a new $p'$ with dead bindings removed, a proof of semantic preservation, and a context morphism from the context of $p$ to the strengthened context of $p'$ with all dead variables removed.

Similarly, the CSE implementation folds over \texttt{Com} recursively, maintaining data structures necessary to map variables and expressions to their canonical form.
At each $(\texttt{let }x = f(v_1, ... v_n)\texttt{ in }b)$ step, we canonicalize the variables $v_i$ to find variables $c_i$.
We then look up the canonicalized expression $f(c_1, \dots, c_n)$ in our data structure to find the canonical variable $c_x$ if it exists and replace $x$ with $c_x$.
If such a canonical $c_x$ does not exist, we add a new entry mapping $f(c_1, \dots, c_n)$ to $x$, thereby canonicalizing any further uses of this expression.

\subsection{Proof Automation for Goal State Simplification in \messa{X}}
\label{sec:automation}

The proof automation tactic $\texttt{simp_peephole } \Gamma$ (used to eliminate framework definitions from the goal state) takes a context $\Gamma$, reduces its type completely, and abstracts out program variables to provide a theorem statement that is universally quantified over the variables of the program, with all framework definitions eliminated.
It uses a set of equation theorems to normalize the type of $\Gamma$.
This is necessary to extract the types of variables during metaprogramming.
Once the type of $\Gamma$ is known, we simplify away all framework definitions (such as \texttt{Expr.denote}).
We then replace all occurrences of a variable accesses $\Gamma[i]$ with a new (Lean, i.e., host) variable.
We do this by abstracting terms of the form $\Gamma[i]$ where $i$ is the $i$-th variable.
This gives us a proof state that is universally quantified over variables from the context.
Finally, we clear the context away to eliminate all references to the context $\Gamma$.
The set of definitions we simplify away is extensible, enabling us to add domain-specific simplification rewrites for the IR.

\section{Pure Rewriting in a Side-Effectful World}
\label{sec:sideeffects}

While \messa{X} streamlines the verification of higher-level IRs that use only value semantics, typical IRs may interleave islands of pure operations (with value semantics) with operations that carry side effects.
An IR that is user-facing can usually be rephrased with high-level, side-effect-free semantics.
Yet, each operation in such an IR is compiled through a sequence of IRs that are lower level and potentially side-effectful.
For example, in the case of FHE, the pure \texttt{FHE} IR is compiled to a lower-level IR that encodes the coset representative of each ideal as an array,
with control flow represented via structured control flow (\texttt{scf}).
Eventually, this is compiled into LLVM which is rife with mutation and global state.
In such a compilation flow, peephole rewrites are used at each intermediate IR to optimize pure fragments while leaving side-effectful fragments untouched.
An effective compiler pipeline introduces the right abstractions to maximize rewrites on side effect-free fragments.

\messa{X} is designed to facilitate verification of peephole rewrites as they arise in such a compiler pipeline.
The previous sections already presented how our framework supports the verification of peephole rewrites in a pure setting.
Yet, our design also allows for the optimization of a pure fragment in a side-effectful context.
We have a mechanized proof of the correctness of the extended framework with support for side effects and a rewrite theorem that performs pure rewrites in the presence of side effects.
The key idea is to annotate each \texttt{Op} with an \texttt{EffectKind}, where \texttt{EffectKind.pure} changes the denotation of the \texttt{Expr} into the \texttt{Id} monad,
while \texttt{EffectKind.impure} denotes into an arbitrary, user-chosen, IR-specific monad.
We also introduce a new notion of monadic evaluation of \texttt{Lets}, which returns a valuation plus a proof that, for every variable $v$ that represents a pure expression $e$ in the sequence of let-bindings, the valuation applied to $v$ agrees with the (pure) denotation of $e$. This proof-carrying definition allows us to use this invariant when reasoning inside a subexpression of a monadic bind.

With the above at hand, the overall rewriter construction and proof strategy remains unchanged,
with the additional constraint of performing rewriting only on those operations marked as \texttt{EffectKind.pure},
and the surrounding monadic ceremony required to show that a pure rewrite indeed does not change the state of pure variables in various lemmas.\footnote{
A limitation of our current mechanization is that we assume that all regions are potentially side-effecting. This is a simplification that shall be addressed in a newer version of the proof.
}

\section{Case Studies}
\label{sec:casestudies}

\newcommand{\FrameworkLoc}{$\approx 2.2k$ LoC}
\newcommand{\BitvectorLoc}{4267}
\newcommand{\FHELoc}{957 LoC}
\newcommand{\ScfLoc}{421 LoC}
\newcommand{\CaseStudiesLoc}{$\approx 5.6k$ LoC}
We mechanize three IRs based on ones found in the MLIR ecosystem as case studies for \messa{X} and show how they benefit from the different aspects of our framework.
Note that the core of our framework (definitions of \texttt{Expr}, \texttt{Com}, \texttt{PeepholeRewrite}, lemmas about these objects, and the peephole rewriting theorem) is \FrameworkLoc{}.
The case studies based on our framework together are \CaseStudiesLoc{}, which stresses the framework to ensure that it scales to realistic formal verification examples.

\subsection{Reasoning About Bitvectors of Arbitrary Width}
\label{sec:bitvectors}

We first demonstrate our ability to reason about a well-established domain of peephole rewrites: LLVM's arithmetic operations over fixed-bitwidth integers.
Using the Z3 SMT solver~\cite{deMoura2008z3}, the Alive project~\cite{lopes2015provably,lopes2021alive2} can efficiently and automatically reason about these.
Notably, at the time of this writing, almost \numalivepatches~LLVM patches have justified their correctness by referencing Alive.
In this way, accessible proof tools can find a place in production compiler development workflows.
However, Alive is limited by the capabilities of the underlying SMT solvers.
SMT solvers are complex, heuristic-driven, and sometimes even have soundness bugs~\cite{DBLP:conf/pldi/WintererZS20}.
They are also specialized to support very concrete theories.
Among others, this means Alive can only reason about a given fixed bitwidth.
Even recent work that specifically aims to generalize rewrites to arbitrary bitwidths, can only exhaustively test a concrete set of bitwidths~\cite{hydra}.
Using our framework, we can reproduce Alive-style correctness proofs, and extend them to reason about arbitrary (universally quantified) bitwidths.
This ability to handle arbitrary bitwidth is important in verification contexts that have wide bitvectors, as they can occur in real-life VLSI problems~\cite{kallstrom2016fast, wang2013novel}.
MLIR itself has multiple IRs that require bitvector reasoning:
\texttt{comb} for combinational logic in circuits, \texttt{arith} and \texttt{index} for integer and pointer manipulation, and \texttt{llvm} which embeds LLVM IR in MLIR.
Our streamlined verification experience offers developers an Alive-style workflow for the \texttt{llvm} dialect, while allowing reasoning across bitwidths.
As our framework is extensible, we believe we can also support other dialects that require bitvector reasoning, such as \texttt{comb}, \texttt{arith}, and \texttt{index}.

\subsubsection{Modeling a fragment of LLVM IR: Syntax and Semantics}

To test our ability to reason about bitvectors in practice, we model the semantics of the arithmetic fragment of LLVM as the IR \messa{\operatorname{LLVM}}.
We support the (scalar) operators:
\texttt{not}, \texttt{and}, \texttt{or}, \texttt{xor}, \texttt{shl}, \texttt{lshr}, \texttt{ashr}, \texttt{urem}, \texttt{srem}, \texttt{add}, \texttt{mul}, \texttt{sub}, \texttt{sdiv}, \texttt{udiv}, \texttt{select} and \texttt{icmp}.
We support all \texttt{icmp} comparison flags, but not the strictness flags \texttt{nsw} and \texttt{nuw}.

At the foundation of our denotational semantics is Lean's \texttt{BitVec} type, which models bitvectors of arbitrary width and offers \texttt{smtlib}~\cite{barrett2010smt} compatible semantics.
However, when we started this work, most bitvector operations were not defined in the Lean ecosystem and the bitvector type itself was not fully fleshed out.
Hence, we worked with the \mathlib and Lean community to build and upstream a theory of bitvectors.\footnote{github.com/leanprover-community/mathlib4/pull/\{5383,5390,5400,5421,5558,5687,5838,5896,7410,\\7451,8231,8241,8301,8306,8328,8345,8353\},\\
github.com/leanprover/lean4/pull/\{3487,3471,3461,3457,3445,3492,3480,3450,3436\}, github.com/leanprover/std4/pull/\{357,359,599,626,633-636,637,639,641,645-648,655,658-660,653\}}
After developing the core theory in \mathlib{}, Lean's mathematical library, development subsequently moved into Lean core, where we continue to evolve Lean's bitvector support.

The semantics of LLVM's arithmetic operations follow the semantics of \texttt{smtlib} (and consequently Lean's) bitvectors closely.
In case of integer wrapping or large shifts, for example, LLVM can produce so-called poison values~\cite{lopes2015provably}, which capture undefined behavior as a special value adjoined to the bitvector domain.
LLVM's poison is designed not to be a side effect and, consequently, can be reasoned about in a pure setting.
In contrast, \texttt{ub} is a side effect that triggers immediate undefined behaviour, and can be refined into any behavior.
In LLVM, the following refinements are legal: $\texttt{ub} \sqsubseteq \texttt{poison} \sqsubseteq \texttt{val}$.
Among the instructions we model, division and remainder can produce immediate undefined behavior $\texttt{ub}$.
In our framework, we approximate these by collapsing the side-effectful undefined behavior and side-effect-free poison both into \texttt{Option.none}.
We thus denote bitvectors into the type \texttt{Option (BitVec w)}.
This is safe as long as the right hand side is allowed to produce \texttt{ub} only when the left hand side produces \texttt{ub}.
In our context, only the division and remainder operations produce \texttt{ub}.
In all the Alive rewrites we translate that contain division and remainder operations, we manually verify that the right hand side of a rewrite triggers \texttt{ub} if and only if the left hand side does
(by checking that any division/reminder on the right has a corresponding operation with syntactically equal arguments on the left).
To be fully correct one can either treat division and remainder as side-effectful operations in our framework or develop further theorey with respect to treating \texttt{ub} as a side effect. We leave separating \texttt{ub} as a side effect distinct from poison, and reasoning about peephole rewrites which refine such side effects as interesting future work.

For side-effect-free programs, our semantics match the LLVM semantics.
We perform exhaustive enumeration tests between our semantics and that of LLVM. 
We take advantage of the fact that an IR with computable semantics automatically defines an interpreter in our framework.
We build an executable program that runs every instruction, with all possible input combinations upto bitwidth $8$.
We get LLVM's ground truth by using LLVM's optimizer, \texttt{opt} to transform the same instruction with constant inputs.
This optimizes the program into a constant output, handling undefined behavior.
By exhaustive enumeration, our tested executable semantics correspond to the LLVM semantics wherever the result is \texttt{Option.some}, and also soundly model undefined behavior whenever the result is \texttt{Option.none}.
This gives us confidence our semantics correspond to LLVM's.

\subsubsection{Proving Bitvector Rewrites in our Framework}


\newcommand{\alivenumaddsub}{19}
\newcommand{\alivenumandorxor}{34}
\newcommand{\alivenumshift}{9}
\newcommand{\alivenumselect}{9}
\newcommand{\alivenummuldivrem}{17}
\newcommand{\alivetotaltranslated}{93}
\newcommand{\alivetotaltestsuite}{184}
\newcommand{\alivetotalinstcombine}{435} 
\newcommand{\alivenumformalized}{1}
\newcommand{\alivenumring}{11}
\newcommand{\alivenumext}{35}

\newcommand{\alivetotalautomated}{54}
\newcommand{\alivetotalhandwritten}{6}

Effective automation for bitvector reasoning is necessary to resolve the proof obligations that \messa{X} derives automatically from peephole rewrites expressed as MLIR program snippets.
While Lean does not yet have extensive automation for bitvectors, thanks to our work we can already use a decision procedure for commutative rings~\cite{DBLP:conf/tphol/GregoireM05} and an extensionality lemma that establishes the equality of bitvectors given equality on an arbitrary bit index.

We test the available automation on a dataset of peephole optimizations from Alive's test suite, consisting of theorems about addition, multiplication, division, bit-shifting and conditionals.
Out of the $\alivetotalinstcombine$ tests in Alive's test suite, we translate $\alivetotaltranslated$ tests which are the ones that are supported by the LLVM fragment we model and without preconditions.
We prove $\alivetotalautomated$ of these rewrites from the Alive test suite automatically.
Some rewrites cannot be handled automatically.
Of those where automation struggles, we manually prove an additional $\alivetotalhandwritten$, selecting the ones where an SMT solver takes long to prove them even for a specific bitwidth (e.g., 64).
Our proofs are over arbitrary (universally quantified) bitwidth, save for some theorems that are only true at particular bitwidths.\footnote{e.g., \texttt{a + b = a xor b} is true only at bitwidth 1.} As an example, let us consider the following rewrites:

\vspace{1mm}
\begin{minipage}{0.4\textwidth}
\begin{leanfootnotesize}

example (w : Nat) :
  [llvm (w)| {
    ^bb0(%X : _, %Y : _):
      %v1 = llvm.sub %X, %X
      %r = llvm.xor %v1, %Y
      llvm.return %r
  }] ⊑ [llvm (w)| {
    ^bb0(%X : _, %Y : _):
      llvm.return %Y
  }] := by
    simp_alive_peephole
    alive_auto

\end{leanfootnotesize}
\end{minipage}
\begin{minipage}{0.4\textwidth}
\begin{leanfootnotesize}

example (w : Nat) :
  [llvm (w)| {
      ^bb0(%X : _, %Y : _):
        %v1 = llvm.and %X, %Y
        %v2 = llvm.or %X, %Y
        %v3 = llvm.add %v1, %v2
        llvm.return %v3
    }] ⊑ [llvm (w)| {
      ^bb0(%X : _, %Y : _):
        %v3 = llvm.add %X, %Y
        llvm.return %v3
    }] := by
      simp_alive_peephole
      <proof omitted>
\end{leanfootnotesize}
\end{minipage}

\vspace{1em}
Note that due to the support of MLIR syntax in our framework, these rewrites are specified in MLIR syntax.
We use a custom extension with the placeholder syntax \texttt{_}, to stand for an arbitrary bitwidth $w$.
Simplification of the framework code with \texttt{simp\_peephole}, yields the proof obligation
\noindent \LightGrayBlueBox{\texttt{(w : Nat) (X Y : BitVec w) ⊢ LLVM.xor (LLVM.sub X X) Y} $\sqsubseteq$ \texttt{Y}}
for the first example.
This proof obligation only concerns the semantics in the semantic domain of bitvectors, it does not feature MLIR and SSA anymore.
This goal is automatically proven by our proof automation for bitvectors, \texttt{alive\_auto}.
The proof for the second example (omitted) is slightly more involved and
currently requires manual intervention. It yields the proof state: \noindent \LightGrayBlueBox{⊢ (B \&\&\& A) + (B ||| A) = B + A},
where the proof follows by reasoning about the addition as a state machine.
In the longer term, we aim to also connect our work to a verified SAT checker that is under development.\footnote{https://github.com/leanprover/leansat}

\subsection{Structured Control Flow}
\label{sec:structured-control-flow}

The examples of IRs we have seen so far are all straight-line code.
In this use case, we show how we can add control flow to existing IRs, thanks to the parametricity of our framework.
We also demonstrate how encoding control flow structures as regions enable succinct proofs for transformations, by exploiting the high-level structure of these operations.
To this end, we model structured control flow as a fragment of the \texttt{scf} IR in MLIR, by giving semantics to two common kinds of control flow:
\texttt{if} conditions and bounded \texttt{for} loops.
Note that we choose to model \emph{bounded} for loops, since these are the loops that are used in MLIR to model high-level operations such as tensor contractions.
A pleasant upshot is that these are guaranteed to terminate, and can thus have a denotation as a Lean function without requiring modelling of nontermination (which is side-effectful).
Our sketch of the extended framework with side effects will be used to pursue this line of research in the future.
The conditionals and bounded for loops allow us to concisely express loop canonicalizations and transformations from MLIR in \messa{\texttt{scf}}.

We built this parametrically over an existing IR $X$ to allow these constructs to be added to an existing IR $X$.
The key idea is that the \texttt{Op} corresponding to \texttt{scf} is parametrized by the \texttt{Op} corresponding to another IR $X$.
Since the only datatypes \texttt{scf} requires are booleans and natural numbers, we ask that the type domain of $X$ contains these types.
We then provide denotations in \messa{scf(X)} for booleans and integers from the type domain of $X$.
Thus, what we encode is $\messa{scf(X)}$, which is an IR for structured control flow parametrized by another, user-defined IR $X$.

\begin{figure}
\begin{leanfootnotesize}
/-- only control flow operations, parametric over another IR Op'  -/
inductive Op (Op': Type) [OpDenote Op' Ty'] : Type
| coe (o : Op') -- coerce Op' to Op
| for (ty : Ty') -- a for loop whose loop carried data is Ty'

instance [I : HasTy Op' Int] : OpSignature (Op Op') Ty' where
  signature
  | .coe o => signature o
  | .for t => ⟨[/-start-/I.ty, /-step-/I.ty, /-niters-/N.ty, /-v-/t],
     /- region arguments: -/ [([/-i-/I.ty, /-v-/t], /-v'-/t)],
     /-return-/t⟩
instance [I : HasTy Op' Int] [OpDenote Op' Ty']: OpDenote (Op Op') Ty' where
  denote
  | .coe o', args', regArgs' => OpDenote.denote o' args'regArgs' -- reuse denotation of o'
  | .for ty, [istart, istep, niter, vstart]ₕ, [f]ₕ =>
	let istart : ℤ := I.denote_eq ▸ istart -- coerce to `int`.
	... -- coerce other arguments
	let loop_fn := ... -- build up the function that's iterated.
	(loop_fn (istart, vstart)).2
\end{leanfootnotesize}
\caption{Simplified implementation of \messa{scf(X)} Observe that the IR is parametrized over another IR \texttt{Op'},
and that we add control flow to the other IR in a modular fashion.}
\label{fig:scf-impl}
\end{figure}

The \texttt{scf.for} operation (\autoref{fig:scf-impl}) has three arguments: the number of times the loop is to be executed, a starting and step value for the iteration, and a seed value for the loop to iterate on.
Note that in the definition, the IR \texttt{Op} is defined parametrically over another IR \texttt{Op'}, and the types of \texttt{Op} are the same as the types of other IR \texttt{Ty'}.
We perform a similar construction for \texttt{if} conditions.

The denotation of the \texttt{for} loop, as well as theorems about loop transformations, follow from \mathlib{}'s theory for iterating functions, \texttt{Nat.iterate}.
The loop body in \texttt{scf.for} has a region that receives the current value of the loop counter and the current iterated value and returns the next iterated value.
We prove the inductive invariant for loops using the standard theory of iterated function compositions ($f^0 = id$, $f^k \circ f^l = f^{k + l}$, $id^k = id$).
We also prove common rewrites over loops: running a \texttt{for} loop for zero iterations is the same as not running a loop at all (dead loop deletion),
two adjacent loops with the same body can be fused into one when the ending index of the first loop is the first index of the second loop (loop fusion),
and a loop whose loop body does not depend on the iteration count can be reversed (loop reversal).
Similarly, we prove that $\texttt{if}~true~e~e' = e$, and $\texttt{if}~false~e~e' = e'$.

These do \emph{not} count as peephole rewrites in our framework, as they are universally quantified over the loop body (which is a region).
This is unsupported --- peephole rewrites in \messa{X} may only have free variables, not free region arguments.

Consider the loop optimization that converts iterated addition into a single multiplication.
Its proof obligation is \LightGrayBlueBox{$(\vdash \lambda x.~x + \delta)^{n}(c) = n\cdot \delta + c$}.
This transformation is challenging to perform in a low-level IR, since there is no syntactic concept of a loop.
However, this transformation \emph{is} a valid peephole rewrite in our framework since it uses a \emph{statically} known loop body.
We showcase how regions permit MLIR (and, consequently, us) to easily encode and reason with commonplace loop transformations.
The parametricity of our framework allows us to prove theorems that are valid on all IR extensions $\texttt{scf}(X)$.

\subsection{Fully Homomorphic Encryption}
\label{sec:fully-homomorphic-encryption}

A key motivation for \messa{X} is to enable specifying formal semantics for high-level, mathematical IRs.
These IRs require access to complex mathematical objects that are available in proof assistants, and verifying rewrites on such IRs is out of practical reach for today's SMT solvers.
As a case study, we formalize the complete ``Poly'' IR.\footnote{as of commit \href{https://github.com/google/heir/tree/2db7701de976f0277f7d3b8be9c65315c647cf79/include/Dialect/Poly}{2db7701de}}
This IR is a work in progress and is in flux, as it is part of the discussion of an upcoming open standard for homomorphic encryption, developed in collaboration by Intel and Google.\footnote{\url{https://homomorphicencryption.org/}}
Contrary to what its naming implies, this IR does \emph{not} model operations on polynomials.\footnote{In the same way that rationals $\mathbb{Q}$ are not pairs of integers $\mathbb{Z} \times \mathbb{Z}$.}
Instead, codewords are encoded as elements in a finitely-presented commutative ring, specifically, the ring $R \equiv (\mathbb{Z}/q\mathbb{Z})[x]/(x^{2^n} + 1)$, where $q, n \in \mathbb{N}$ are positive integers ($q$ composite).
The name ``Poly'' comes from the equivalence class representatives are polynomials, but not all IR operations are invariants of the equivalence class.

The ``Poly'' IR is, in fact, a superset of the \texttt{QuotRing} IR we defined in \autoref{sec:mlir-intro}.
It consists of the operations \texttt{add}, \texttt{sub}, \texttt{mul}, \texttt{mul\_constant}, \texttt{leading_term}, \texttt{monomial}, \texttt{monomial_mul}, \texttt{from\_tensor}, \texttt{to\_tensor}, \texttt{arith.constant} and \texttt{constant}.\footnote{It also has distinct types for integers and naturals, which we unified in \autoref{sec:mlir-intro} for simplicity.}

Most of these operations are self-explanatory and derive from the (commutative) ring structure of $R$ or are used to build elements in $R$, like the equivalence classes of constants or monomials.
Three operations, \texttt{to\_tensor} and \texttt{from\_tensor} and \texttt{leading_term} do not follow directly from the algebraic properties of the polynomial ring.
Instead, they depend on a (non-canonical) choice of representatives for each ideal coset in the polynomial ring.
More precisely, let $\pi: (\mathbb{Z}/q\mathbb{Z})[x] \twoheadrightarrow (\mathbb{Z}/q\mathbb{Z})[x]/(x^{2^n} + 1)$ be the canonical surjection into the quotient, taking a polynomial to its equivalence class modulo division by $x^{2^{n}} + 1$.
Further let $\sigma: (\mathbb{Z}/q\mathbb{Z})[x]/(x^{2^n} + 1) \hookrightarrow \mathbb{Z}/q\mathbb{Z}[x]$ be the injection taking an equivalence class to its (unique) representative with degree $\leq 2^{n}$.
This is a right-inverse of $\pi$, i.e. $\pi \circ \sigma = id$.
Note that multiple right-inverses could have been chosen for $\sigma$:
As long as $\sigma(x)$ is a representative of the equivalence class of $x$ for all $x\in (\mathbb{Z}/q\mathbb{Z})[x]/(x^{2^n} + 1)$, $\sigma$ will be a right-inverse of $\pi$.
The operation \texttt{to\_tensor(p)} returns the vector $(\sigma(p)[i])_{i=0,\ldots,2^n}$, where $a[i]$ represents the i-th coefficient, i.e. $\sigma(p) = \sum_{i=0}^{2^{n}} (\sigma(p)[i]) x^{i}$, and \texttt{to\_tensor} the converse.
Similarly, \texttt{leading\_term(p)} returns the equivalence class of the leading term of the representative $\sigma(p)$ (which also depends on the choice of $\sigma$).

This allows us to define the semantics and prototype both the IR and rewrites in it.
Rewrites like \texttt{mul(p,q)} $\rightarrow$ \texttt{mul(q,p)} follow immediately from the fact that $R$ is a commutative ring.
Other rewrites like \texttt{from\_tensor(to\_tensor(p))} $\rightarrow$ \texttt{p}, or even \texttt{add(p,monomial(1,$2^n$))} $\rightarrow$ \texttt{sub(p,1)}, on the other hand, are more specific to this IR and have a higher manual-proof overhead.
We prove all of these.

We discussed the IR and potential semantics with the authors of the HEIR IR in the context of the upcoming open standard for homomorphic encryption.
We believe that a framework like the one presented in this paper will allow standards like these to be defined with formal semantics from the ground up.

\section{Related Work}
\label{sec:related}

The semantics of LLVM, the spiritual ancestor of MLIR, have been well-studied.
Both Vellvm~\cite{zhao2012formalizing} and K-LLVM~\cite{li2020k} formalized a large portion of LLVM, including reasoning about SSA transformations explicitly~\cite{zhao2013formal}.
Alive~\cite{lopes2015provably} and Alive~2~\cite{lopes2021alive2} provide push-button verification for a subset of LLVM by leveraging SMT solvers.
Alive-tv does the same for a set of concrete IRs for tensor operations in MLIR~\cite{DBLP:conf/cav/BangNCJL22}.
AliveInLean \cite{lee2019aliveinlean} proves the correctness of the translation from the Alive DSL into SMT expressions,
as well as the correctness of their encoding of program refinement as an SMT expression.
In contrast, our work focuses on building a framework for describing full programs (rather than rewrite snippets),
and formally defines and proves the correctness of peephole rewriting within a larger program context.
The semantics and correctness of compiling compositionally have been explored by multiple authors, like Pilsener~\cite{neis2015pilsner} or many variants of CompCert~\cite{leroy2016compcert}:
like compositional CompCert~\cite{stewart2015compositional}, CompCertX~\cite{wang2019abstract}, SepCompCert~\cite{kang2016lightweight}, CompCertM~\cite{song2019compcertm}, and CompCertO~\cite{koenig2021compcerto}.
A great summary of the approaches to this problem (including the ones mentioned above), with their differences and similarities, is given by Patterson et al~\cite{patterson2019next}.
All of these use fixed languages but are reasonable ways of giving semantics to relevant IRs in \messa{X}.

The authors of~\cite{zakowski2021modular} introduce a more modular approach to LLVM's semantics, based on interaction trees~\cite{xia2019interaction}.
Like theirs, our semantics is also denotational and can be executed.
We currently only model a restricted set of side effects, whereas interaction trees shine when modeling more complex side effects such as memory or non-terminating behavior.
An approach like this would thus be a great candidate for the semantics of a lower-level IR such as LLVM within MLIR.
Similarly, the work of DimSum~\cite{sammler2023dimsum} deals with the boundaries between languages in the context of linking.
This addresses also important part aspect we don't model yet: what occurs at the boundaries of IRs, when mixing them.

There is a longer line of work studying SSA and its relationship to functional programming. Our work is inspired by and builds on the ideas from~\cite{kelsey1995correspondence,appel1998ssa,barthe2014formal}.
Complex compiler optimizations have also been studied formally and verified, like~\cite{courant2021verified} which implements verified polyhedral optimization.
We focus on the simpler and more ubiquitous peephole rewrites.

Our work differs from prior work on formalizing peephole rewrites by providing a framework for reasoning about SSA peephole rewrites.
The closest similar work, Peek~\cite{mullen2016verified} defines peephole rewriting over an assembly instruction set.
Their rewriter expects instructions to be adjacent to one another.
Furthermore, their rewriter restricts source and target patterns to be of the same length, filling in the different lengths with \texttt{nop} instructions.
Their patterns permit side effects, which we disallow since we are interested in higher-level, pure rewrites.
Our patterns provide more flexibility since the source and target patterns are arbitrary programs, and are matched on sub-DAGs instead of a linear sequence.

\section{Conclusion}
\label{sec:conclusion}

Peephole rewrites represent a large and important class of compiler optimizations.
We have seen how domain-specific IRs in SSA with regions greatly extend the scope of these peephole rewrites.
They raise the level of abstraction both syntactically with def-use chains and nesting, and semantically, with domain-specific abstractions.
We have shown how to reason effectively about such SSA-based compilers, and, specifically, local reasoning in the form of peephole rewrites.
We advocate building on top of a proof assistant with a small TCB, an expressive language and a large library of mathematics.
This increases the confidence in our verification and extends its applicability to many domains where more specialized methods don't exist.
We also advocate proof automation and an intrinsically well-typed mechanized core that can be designed to focus on the semantics of the domain.
We incarnate these principles in \messa{X}, a framework built on Lean and \mathlib to reason about domain-specific IRs in SSA with regions.
We show how \messa{X} is simple to use, amenable to automation, and effective for verifying IRs over complex domains.

\bibliography{references}

\end{document}